\documentclass[twocolumn]{aastex631}

\shortauthors{Nakata et al.}

\usepackage{color}
\usepackage{amsmath}
\usepackage{bm}
\usepackage{enumitem}

\let\Re\relax
\DeclareMathOperator{\Re}{Re}

\begin{document}

\title{The Simons Observatory: Detector Polarization Angle Calibration using Sparse Wire Grid \\
with Initial Data Sets of the Small Aperture Telescope}

% First Tier
\author[0000-0002-6300-1495]{Hironobu Nakata}
\affiliation{Department of Physics, Faculty of Science, Kyoto University, Kyoto 606-8502, Japan}

\author[0000-0002-0400-7555]{Shunsuke Adachi}
\affiliation{Okayama University, Department of Physics, Okayama 700-8530, Japan}

\author[0000-0003-0221-2130]{Kyohei Yamada}
\affiliation{Joseph Henry Laboratories of Physics, Jadwin Hall, Princeton University, Princeton, NJ 08544, USA}

\author[0009-0009-9806-2317]{Michael Randall}
\affiliation{Department of Astronomy \& Astrophysics, University of California San Diego, San Diego, CA, USA}

\author[0009-0001-3477-5141]{Yutaro Kasai}
\affiliation{Department of Physics, Faculty of Science, Kyoto University, Kyoto 606-8502, Japan}

% Alphabetic Tier
\author[0000-0002-3407-5305]{Kam Arnold}
\affiliation{Department of Astronomy \& Astrophysics, University of California San Diego, San Diego, CA, USA}
\affiliation{Department of Physics, University of California San Diego, San Diego, CA, USA}

\author[0009-0008-4312-6814]{Bryce Bixler}
\affiliation{Department of Physics, University of California San Diego, San Diego, CA, USA}

\author[0000-0002-3266-857X]{Yuji Chinone}
\affiliation{QUP (WPI), KEK, Tsukuba, Ibaraki 305-0801, Japan}
\affiliation{Kavli IPMU (WPI), UTIAS, The University of Tokyo, Kashiwa, Chiba 277-8583, Japan}

\author[0000-0001-5068-1295]{Kevin T. Crowley}
\affiliation{Department of Astronomy \& Astrophysics, University of California San Diego, La Jolla, CA 92093 USA}

\author[0009-0006-7382-1434]{Nadia Dachlythra}
\affiliation{Department of Physics, University of Milano-Bicocca, Piazza della Scienza 3, 20126, Milano, Italy}

\author[0009-0003-5814-2087]{Samuel Day-Weiss}
\affiliation{Joseph Henry Laboratories of Physics, Jadwin Hall, Princeton University, Princeton, NJ 08544, USA}

\author[0000-0001-7225-6679]{Nicholas Galitzki}
\affiliation{Department of Physics, University of Texas at Austin, Austin, TX, 78712, USA}
\affiliation{Weinberg Institute for Theoretical Physics, Texas Center for Cosmology and Astroparticle Physics, Austin, TX 78712, USA}

\author[0000-0002-8340-3715]{Serena Giardiello}
\affiliation{School of Physics and Astronomy, Cardiff University, UK}

\author[0000-0002-6898-8938]{Bradley R. Johnson}
\affiliation{University of Virginia, Department of Astronomy, Charlottesville, VA 22904, USA}

\author[0000-0003-3118-5514]{Brian Keating}
\affiliation{UC San Diego Dept of Physics, 9500 Gilman Dr, La Jolla, CA 92093 USA}

\author[0000-0003-0744-2808]{Brian J. Koopman}
\affiliation{Wright Laboratory, Department of Physics, Yale University, New Haven, Connecticut 06511, USA}

\author[0009-0004-9631-2451]{Akito Kusaka}
\affiliation{Department of Physics, The University of Tokyo, Tokyo 113-0033, Japan}
\affiliation{Research Center for the Early Universe, School of Science, The University of Tokyo, Tokyo 113-0033, Japan}
\affiliation{Kavli Institute for the Physics and Mathematics of the Universe (WPI), UTIAS, The University of Tokyo, Kashiwa, Chiba, 277-8583, Japan}
\affiliation{Physics Division, Lawrence Berkeley National Laboratory, Berkeley, CA 94720, USA}

\author[0000-0002-6522-6284]{Jack Lashner}
\affiliation{Wright Laboratory, Department of Physics, Yale University, New Haven, Connecticut 06511}

\author[0000-0002-8307-5088]{Federico Nati}
\affiliation{Department of Physics, University of Milano-Bicocca, Piazza della Scienza 3, 20126 Milano (MI), Italy}

\author[0000-0002-9828-3525]{Lyman Page}
\affiliation{Joseph Henry Laboratories of Physics, Jadwin Hall, Princeton University, Princeton, NJ 08544, USA}

\author[0009-0003-2513-2608]{Daichi Sasaki}
\affiliation{Department of Physics, The University of Tokyo, Tokyo 113-0033, Japan}

\author[0000-0002-3644-2009]{Yoshinori Sueno}
\affiliation{Joseph Henry Laboratories of Physics, Jadwin Hall, Princeton University, Princeton, NJ 08544, USA}

\author[0000-0001-6816-8123]{Junya Suzuki}
\affiliation{Department of Physics, Faculty of Science, Kyoto University, Kyoto 606-8502, Japan}

\author[0000-0003-2439-2611]{Osamu Tajima}
\affiliation{Department of Physics, Faculty of Science, Kyoto University, Kyoto 606-8502, Japan}
\affiliation{Kavli IPMU (WPI), UTIAS, The University of Tokyo, Kashiwa, Chiba 277-8583, Japan}
\affiliation{High Energy Accelerator Research Organization (KEK), Tsukuba, 305-0801, Japan}

\author[0000-0002-1667-2544]{Tran Tsan}
\affiliation{Physics Division, Lawrence Berkeley National Laboratory, Berkeley, CA 94720, USA}

\begin{abstract}

Improved measurements of $B$-modes in the cosmic microwave background can be obtained through accurate calibration of the orientation of detector antennas as projected onto the sky.
Miscalibration of the detector polarization angle leads to a leakage of $E$-modes into $B$-modes, which can bias the detection of the latter.
To achieve a $\sigma(r)$ of 0.003, the Simons Observatory Small Aperture Telescopes are required to calibrate the global polarization angle on the sky with an accuracy ${\lesssim}0.1^\circ$.
We demonstrate a fully remote-controllable calibration system using a ``sparse wire grid,"
which injects a rotatable linear polarized signal across the telescope's focal plane.
This calibration system is installed and operational on a Small Aperture Telescope at its observing site at the Parque Astron\'omico in the Atacama desert in Chile.
We developed a pipeline for the detector polarization angle calibration, and demonstrate it using initial data for 93~GHz and 145~GHz frequency bands.
The observed distribution of detector polarization angles is in agreement with the instrument design.
Statistical uncertainties for the relatively calibrated polarization angles are $0.02^\circ$ and $0.03^\circ$ at 93~GHz and 145~GHz, respectively.
Systematic uncertainty was evaluated to be $0.08^\circ$ at the hardware development and fabrication stage.
Their sum in quadrature is less than $0.1^\circ$.

\end{abstract}


\section{Introduction} \label{sec:intro}

\begin{figure*}[t!]
    \centering
    \includegraphics[width=1.4\columnwidth]{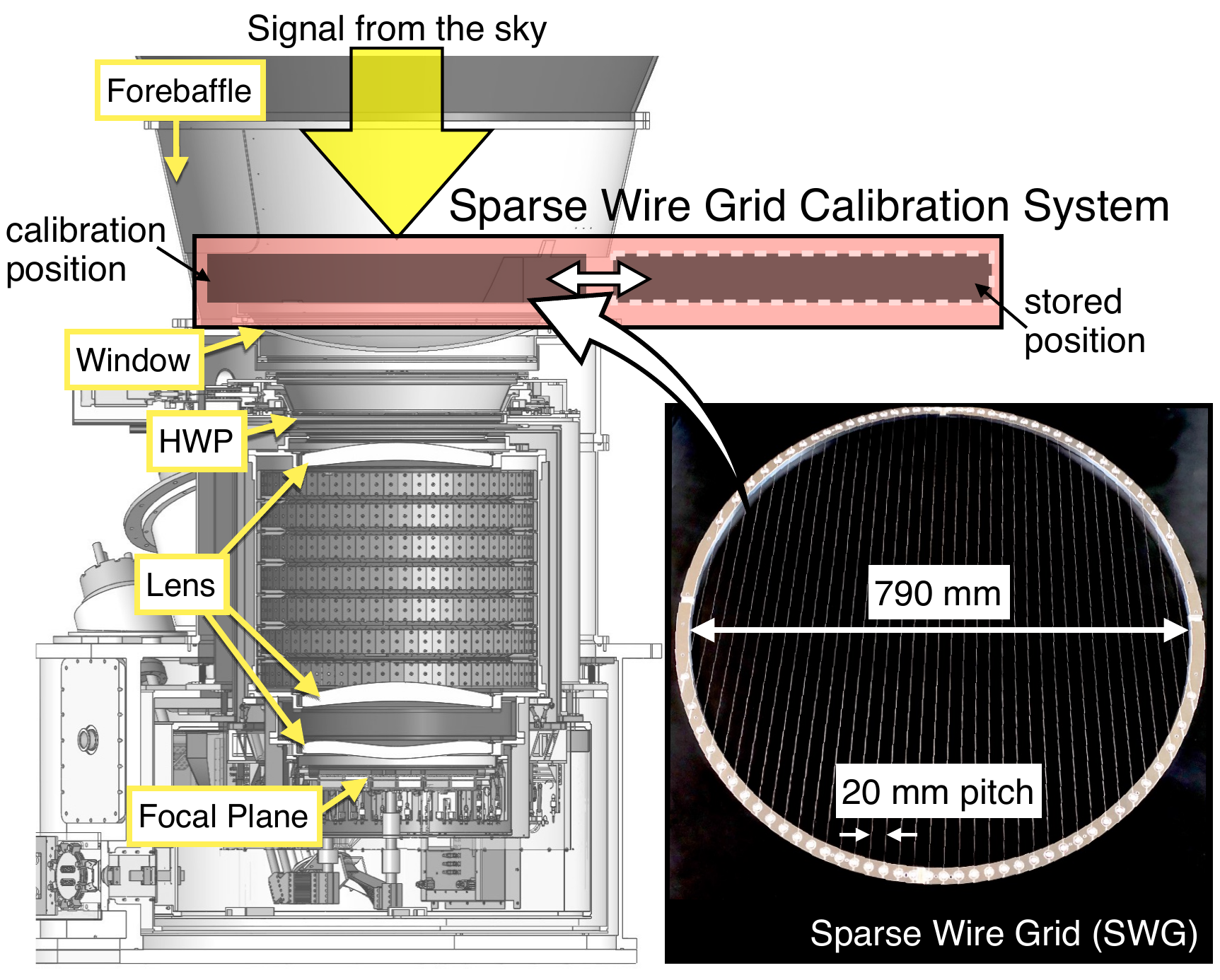}
    \caption{
        A cross-section of a small aperture telescope (SAT) and the sparse wire grid (SWG) calibration system.
        A photo of the SWG is also shown.
        The big yellow arrow from the top indicates the incident light from the sky.
        The position of the SWG is remotely set at ``calibration position'' or ``stored position'' by using two linear actuators.
        Detectors on the focal plane receive signals passed through the optical elements, and the SWG is the first element viewed from the sky side.
        }
    \label{fig:sat_cross_sec}
\end{figure*}

The cosmic microwave background (CMB) is a powerful scientific probe for unveiling the early universe.
In particular, precise measurements of the CMB polarization patterns could allow us to detect the signature of the primordial gravitational wave background, which is predicted by cosmic inflation \citep{Starobinsky_1979, Basko_Polnarev_1980, Polnarev_1985,Kamionkowski_1997, Seljak_1997}.
The amplitude of the primordial gravitational wave background is given by $r$, the ratio of the amplitudes of the tensor and scalar primordial perturbations.
The CMB polarization patterns are parametrized by a parity-even component ($E$-mode) and a parity-odd component ($B$-mode).
The primordial $B$-mode can be sourced by the primordial tensor perturbations, while the $E$-mode are produced by both scalar and tensor perturbations.
Although the \textit{E}-mode spectrum has been precisely measured by previous CMB experiments such as Planck satellite \citep{Planck2018resultsV}, ACT \citep{Louis_2025}, and SPTpol \citep{SPTpol2025},
the quest for the primordial \textit{B}-mode continues.
The current upper limit is $r < 0.032$ at $95\%$ confidence level \citep{BICEPPlanck2022}.
Next-generation experiments are designed to measure CMB polarization with higher sensitivity and get more stringent constraints on $r$.

The Simons Observatory (SO) experiment is designed to measure the CMB intensity and polarization at an observing site in the Atacama Plateau at an altitude of 5,200 m in Chile.
The Small Aperture Telescopes (SATs) aim to detect or limit the primordial gravitational wave signature at the level of $\sigma(r) = 0.003$ \citep{SOScienceForecast} with three SATs (for $r=0$).
Figure \ref{fig:sat_cross_sec} shows a cross-section of the SAT.
The incident light illuminates the focal plane through a window, continuously-rotating half-wave plate (HWP), and three silicon lenses \citep{Galitzki_2024}.
The HWP modulates a linearly polarized signal, which can be recovered through demodulation of the detector timesreams \citep{ABS_HWP, SOHWPDesign}.
The focal plane is designed with 12,040 optically coupled transition edge sensor (TES) bolometers distributed across seven hexagonal detector modules and split evenly between the 93~GHz and 145~GHz frequency bands.
Each module for the SAT is designed to have 860 TES bolometers for the 93~GHz band and 860 TES bolometers for the 145~GHz band.
The goal of $\sigma(r) = 0.003$ requires control over contaminants such as instrumental systematics and astrophysical foregrounds. In this work we focus on the calibration of the polarization sensitive orientation of the detectors.
Calibration of the detector polarization angle (i.e., understanding the antenna orientation relative to an incoming polarization signal) is important because the miscalibration of the detector polarization angle leads to a leakage of the \textit{E}-mode into the \textit{B}-mode \citep{Bunn_2003, Hu_2003}.
Leakage from $\mathcal{O}{(0.5^\circ)}$ miscalibration is approximately equivalent to $r = 0.01$ \citep{Bryan_2018}.
Our requirement for the overall accuracy of the detector polarization angle is ${\lesssim}0.1^\circ$ for the primordial $B$-mode analysis \citep{Bryan_2018, Abitbol_2021}.
The current accuracy of the polarization angle calibration using astronomical sources, such as the Crab Nebula (also called Tau\,A), is $0.33^\circ$ \citep{TauAPolarization}.

An artificial calibration system is essential to establish a more accurate calibration standard.
In SO, there are two approaches for the artificial calibration system: a aperture-filling calibrator using a sparse wire grid (SWG) \citep{Tajima_2012, SOWireGridDesign} and a far field calibrator using a drone \citep{Federico_2017, Coppi_2025}.

The SWG is an artificial polarized source consisting of parallel metal wires that reflect ambient thermal radiation into the receiver.
The hardware system of the SWG measures the orientation of the detector antenna relative to the linearly polarized light from the wires.
The direction of this polarized light is aligned with the physical orientation of the SWG.
Therefore, given the SWG's orientation to a coordinate system, we can calibrate the detector antenna's orientation in that coordinate system.
In this study, we demonstrate the methodology to calibrate the detector polarization angle using initial data from the SWG on one of  the SATs, which internally called SATp1 (hereafter referred to as the SAT).
In Section \ref{sec:sparse_wire_grid}, we describe the SWG calibration system and its operation procedure.
Methodology of the data analysis, from time-ordered data (TOD) to the detector polarization angle, is described in Section \ref{sec:method}.
The calibration results are shown and discussed in Section \ref{sec:evaluation}.
Our conclusions are given in Section \ref{sec:conclusion}.

\section{Instrument and Data} \label{sec:sparse_wire_grid}

The mechanical details of the calibration system are described in \citet{SOWireGridDesign}.
The SWG is set outside the window as shown in Figure~\ref{fig:sat_cross_sec}.
We can move it in and out of the primary beam of the telescope and change the SWG's angle with respect to the focal plane remotely.
When we calibrate the detector polarization angle, we move it into the beam (``calibration position''),
and when we observe the CMB, we move it out of the way and store it at the side of the telescope (``stored position'').
Four limit switches incorporated in the edges of the two linear actuators detect the position of the SWG.
During the calibration, the telescope is stationary.

The SWG consists of 39 parallel tungsten wires with a spacing of 20~mm on an aluminum frame (see a photo in Figure~\ref{fig:sat_cross_sec}).
The diameter of each wire is 0.1~mm.
The deviation of the wires from parallel is less than $0.02^\circ$, which is guaranteed by the manufacturing tolerance.

\begin{figure}[t!]
    \centering
    \includegraphics[width=0.8\columnwidth]{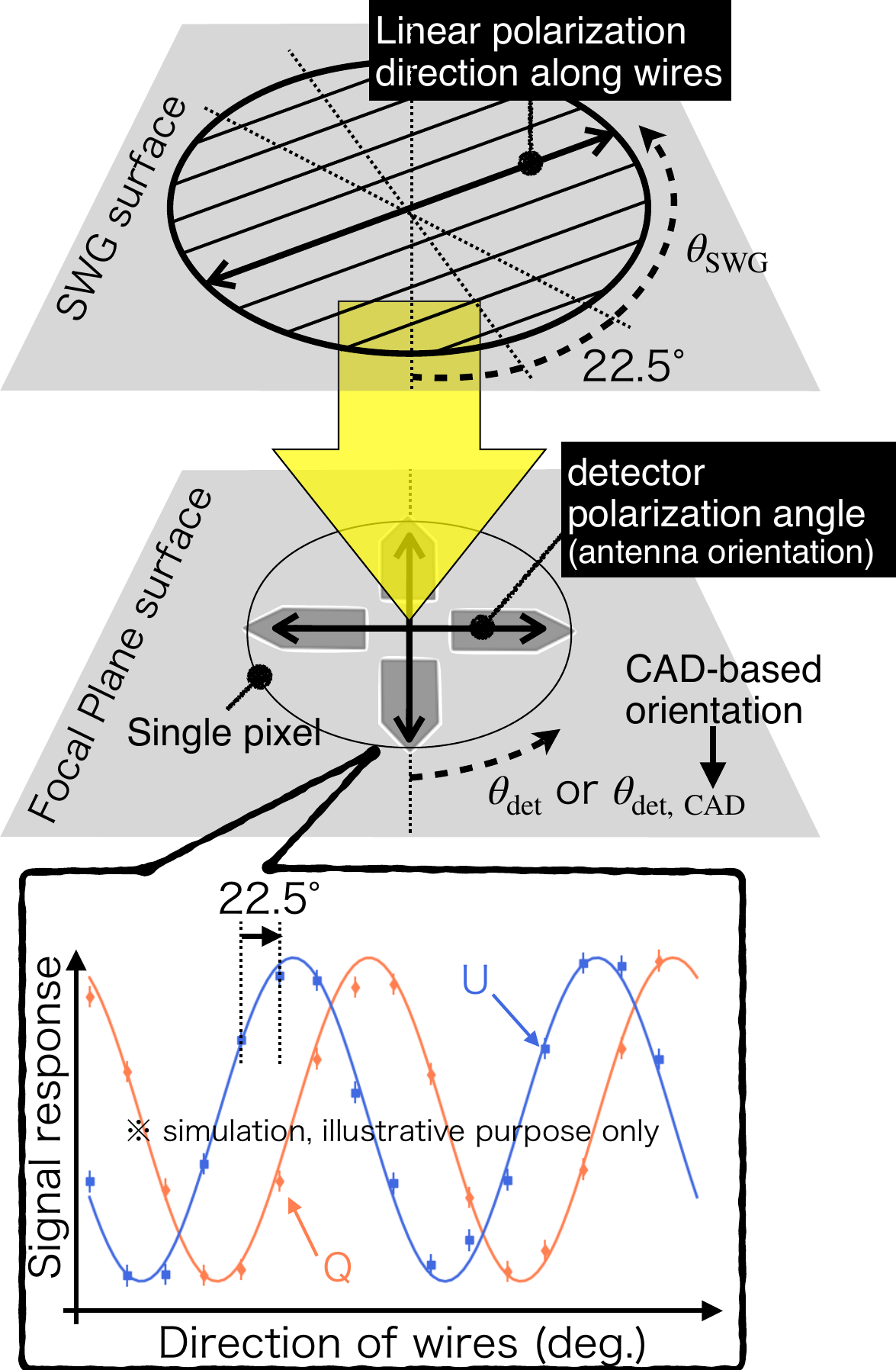}
    \caption{
        An illustration of the SWG calibration.
        The positive direction of the SWG rotation, $\theta_\mathrm{SWG}$, is defined as counterclockwise viewed from above the telescope.
        The operation of the SWG is described in the text.
        A single detector pixel on the focal plane has two orthogonal antenna orientations.
        We use $\theta_\mathrm{det}$ as the calibrated polarization sensitive direction of the detector, and $\theta_\mathrm{det,\,CAD}$ as the design angle specified in the computer aided design (CAD) of the detector module.
        The simulated response of a single detector is also shown.
        The response in stokes parameters is obtained after the demodulation (see Section~\ref{subsec:demodulation}).
    }
    \label{fig:swg_signal}
\end{figure}

The SWG reflects ambient radiation, primarily originating from the forebaffle, and the reflected radiation is linearly polarized in the direction of the wires.
As illustrated in Figure~\ref{fig:swg_signal}, we are able to change the polarized direction of the SWG with respect to the detector.
We extract the detector polarization angles based on the measured responses at each SWG rotation angle.
This procedure also allows us to distinguish between the SWG signal and other polarized sources, such as for reflected radiation at clouds \citep{Takakura_2019, Li_2023, Coerver_2025}. We expect such polarized signals to be negligible provided nominal weather conditions.
While our calibration method enables separation of polarization contamination, we measure polarization residuals consistent with zero in our first calibration run (see Section~\ref{subsec:calibration}).

\begin{figure}[t!]
    \centering
    \includegraphics[width=0.99\columnwidth]{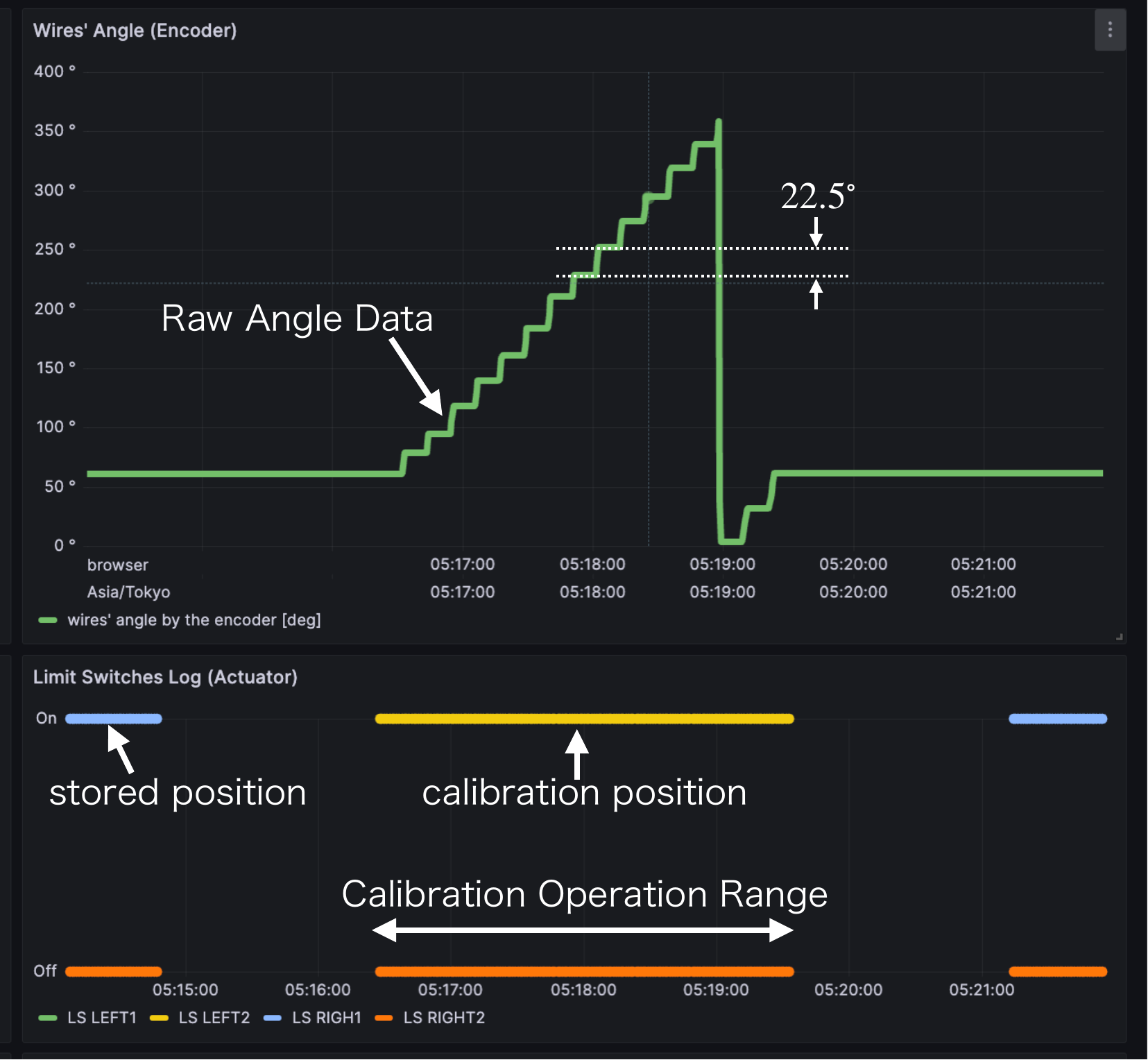}
    \caption{
        An online status monitor of the SWG.
        The raw data encoding wire direction is shown with the green line in the upper panel, tracking the stepwise rotation of the SWG.
        The spatial intervals between the measurement angles are approximately 22.5 degrees.
        The operation range is defined by four limit switches (LS) highlighted as solid lines in the bottom panel.
    }
    \label{fig:grafana_status}
\end{figure}

A single calibration run consists of three steps:
\begin{enumerate}[itemsep=2pt, parsep=0pt, topsep=2pt]
    \item Insertion of the SWG into the calibration position,
    \item Stepwise rotation of the SWG at 16 spaced measurement angles of approximately $22.5^\circ$,
    \item Removal of the SWG from the calibration position.
\end{enumerate}
In the second step, the SWG stays for approximately 10 seconds at each measurement angle, $\theta_\mathrm{SWG} = 22.5^\circ$, ..., $360^\circ$.
The measurement angles are recorded by a magnetic encoder attached to the SWG with a resolution of $0.002^\circ$.
The detector TOD is continuously recorded during the calibration run.
The stepwise approach relaxes the requirements on timing synchronization between the SWG motion and the detector readout.
Figure~\ref{fig:grafana_status} shows an online status panel during a single calibration run.
A single calibration run typically takes 10 minutes.
The SWG rotates only while the limit switch nearby the window is turned on.

In this study, three data sets were taken during the initial science observation period of the SAT.
The first and second data sets were taken on September 7th, 2024, and the third data set was taken on June 7th, 2024.
The precipitable water vapor during all three data sets was $0.5{\sim}0.6$~mm, ensuring similar atmospheric conditions. For both the SWG and the HWP, counterclockwise rotation as viewed from the sky side is defined as the positive direction, and clockwise rotation is defined as negative direction.
In the first and second data set the HWP rotation frequency, $f$, was set to $+2~\mathrm{Hz}$ and $-2~\mathrm{Hz}$, respectively. In the third data set, the SWG was kept stationary and the HWP rotation speed was changed from $+2~\mathrm{Hz}$ to $-2~\mathrm{Hz}$.
To determine the overall rotation of the SWG relative to the sky coordinate, the gravity reference is obtained with a tilt sensor equipped with the SWG system \citep{SOWireGridDesign}.
Data from the tilt sensor were not available due to equipment failure.
Thus, in the following analysis, the overall angle offset to the global polarization angle -- common to all detectors relative to the sky coordinate -- is taken as arbitrary.
In addition, we evaluated only the central detector module data in this study because the validation of the calibration method is the most robust with normal incident radiation.

\section{Data Analysis} \label{sec:method}

\subsection{Data selection} \label{subsec:data_selection}

At first, we select detectors biased in their transition edge \citep{Irwin2005}.
We required the resistance value to be $5\%\text{--}95\%$ of the normal resistance, and the saturation power to be in the range of 1--20~pW.
These are the criteria to remove latched or saturated detectors.
About 450 (430) detectors at 93 (145)~GHz on the central detector module passed these criteria.
These values correspond to approximately $50\%$ yield of the total detectors on the module \footnote{This yield is not a typical value for the CMB observation.}.

\subsection{Demodulation} \label{subsec:demodulation}

The linearly polarized calibration signal produced by the SWG is modulated by the continuous rotation of the HWP.
The modulated data (i.e., time-ordered-data (TOD)) for a single detector at the $j$-th measurement angle  of the SWG, denoted as $d_\mathrm{mod}^{(j)}(t)$, and which is expressed by using Mueller formalism, is:
\begin{equation}
    d^{(j)}_\mathrm{mod}(t) = I + \Re{\left[\left(A_\mathrm{SWG} \exp{2i\theta^{(j)}_\mathrm{SWG}} + \bm{\varepsilon}\right) \exp{2i\Theta}\right]},
    \label{eq:tod}
\end{equation}
where the first term $I$ is the total power, the second term is the polarized component,
$A_\mathrm{SWG}$ is the intensity of the polarization signal from the SWG ($A_\mathrm{SWG} {\sim}1~\mathrm{K}$ in typical conditions) and $\theta^{(j)}_\mathrm{SWG}$ is the SWG direction at $j$-th measurement angle.
The $\bm{\varepsilon}$ is a residual offset.
For simplicity, we ignore the modulation efficiency of the HWP, which is approximately $0.99$ \citep{SOHWPDesign}.
The phase $\Theta$ in the polarized term is
\begin{equation}
    \Theta \equiv -2~\omega(t - \tau_\mathrm{det}) + \theta_\mathrm{det},
    \label{eq:theta}
\end{equation}
where $\omega$ is the angular velocity of the HWP rotation ($\omega = 2\pi f$), and $\theta_\mathrm{det}$ is the detector polarization angle.
Each detector has finite time delay in the response, which is called time constant, $\tau_\mathrm{det}$ \citep{Irwin2005}.
The time constant shifts the detector phase by $2\omega\tau_\mathrm{det}$, typically ${\sim} 1.4^\circ$ with $\tau_\mathrm{det} {\sim} 1~\mathrm{msec}$ and $f = 2~\mathrm{Hz}$.
In Section~\ref{subsec:calibration}, we extract the detector polarization angle by ignoring the angle shift due to $\tau_\mathrm{det}$.
Then, we correct this angle shift in Section~\ref{subsec:tc}.

\begin{figure}[t!]
    \centering
    \includegraphics[width=0.95\columnwidth]{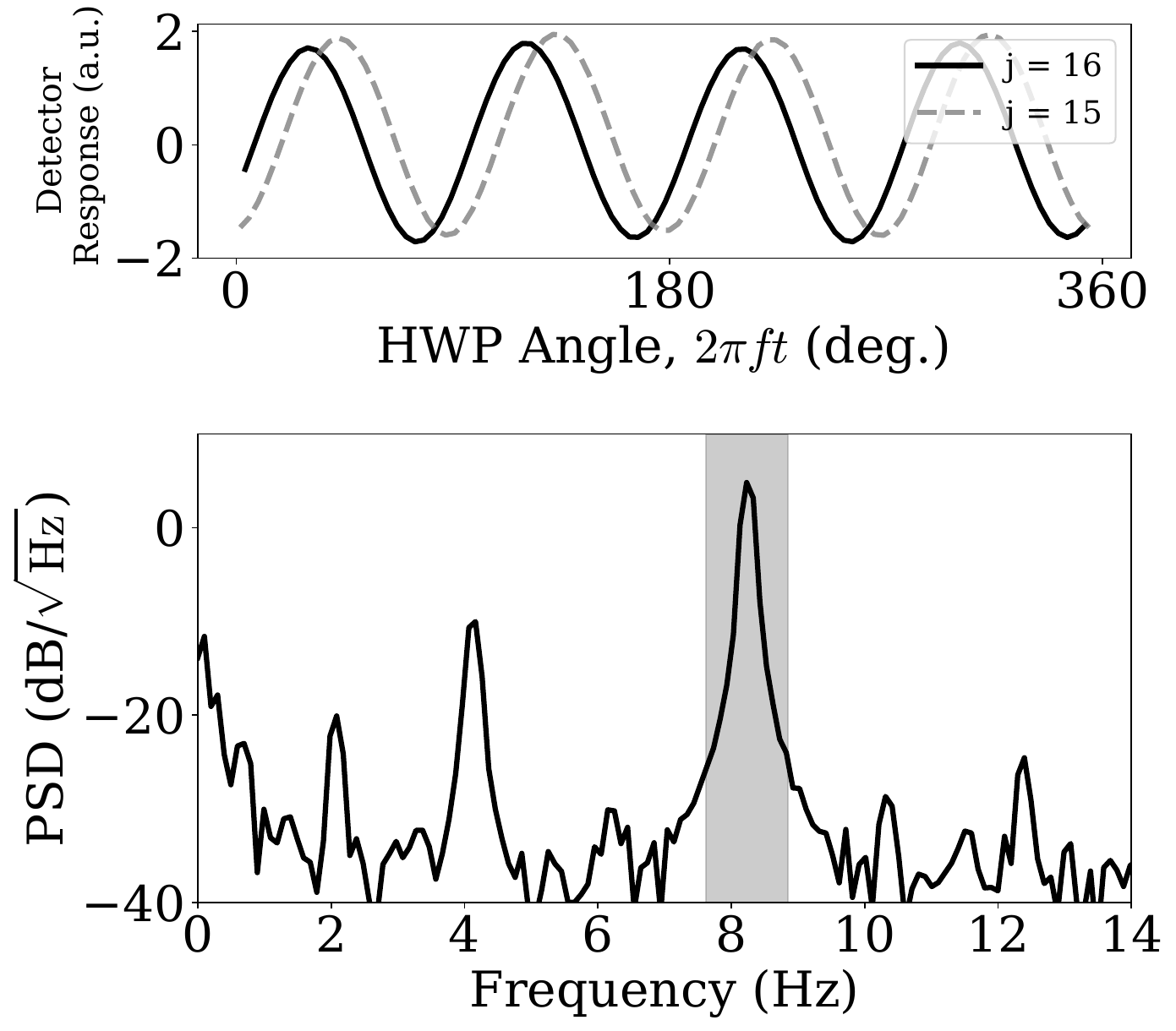}
    \caption{
        (Top panel) The modulated signal response of a single detector (in arbitrary units, a.u.)  as a function of the HWP angle.
        The solid line and the dashed line are the response at the 16th and 15th measurement angles of the SWG rotation, respectively.
        Their phase difference represents the difference of the SWG direction, $22.5^\circ$.
        (Bottom panel) The power spectral density (PSD) of the modulated signal.
        The peaks at around 2, 4 and 8 \rm{Hz} correspond to the $1f$, $2f$, and $4f$ harmonics of the HWP rotation.
        The shaded range is defined as the signal range to be demodulated with a band-pass filter.
        Details are described in the text.
    }
    \label{fig:calibration_operation}
\end{figure}

\begin{figure}[t!]
    \centering
    \includegraphics[width=0.9\columnwidth]{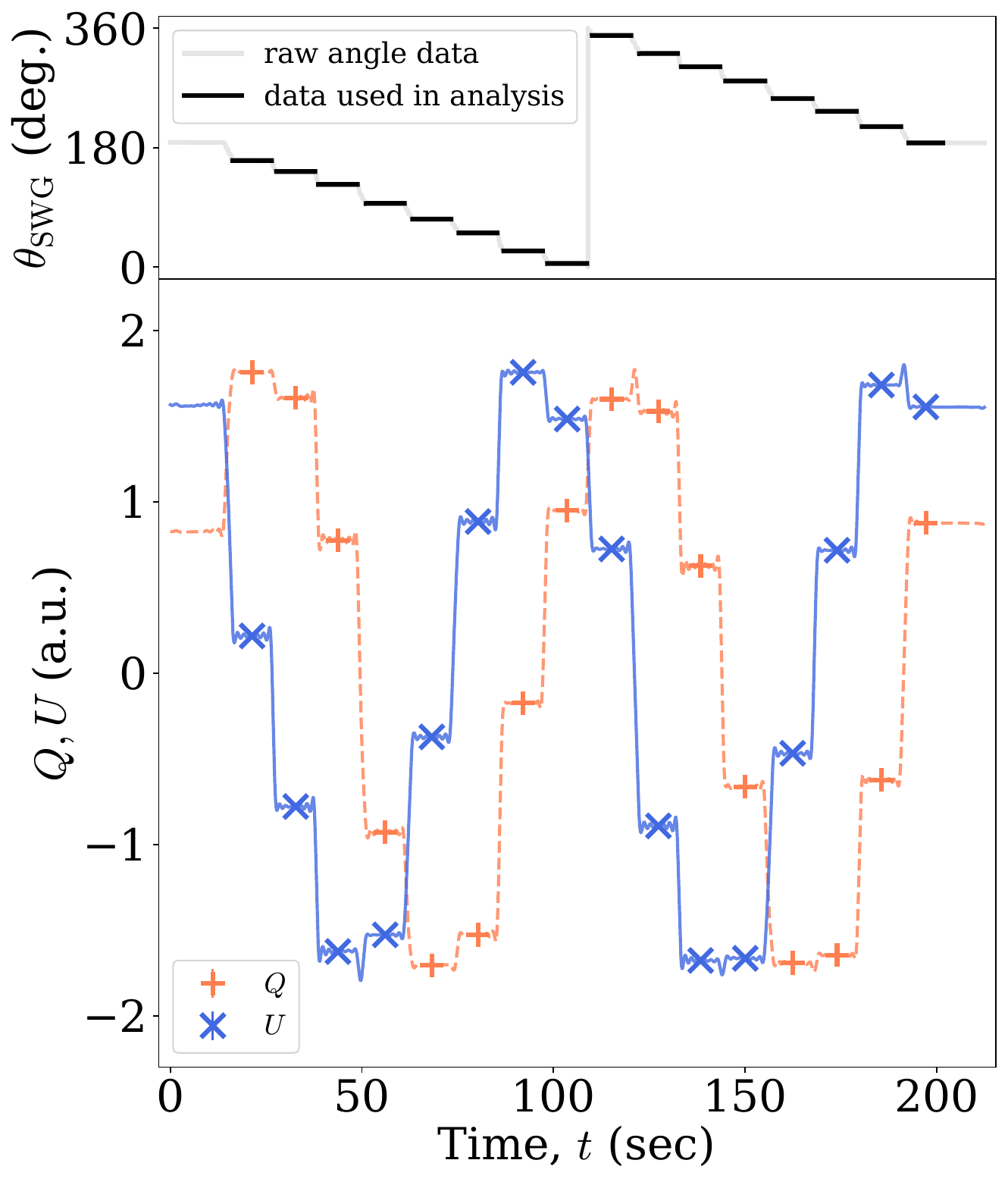}
    \caption{
        (Top panel) The direction of the SWG, $\theta_\mathrm{SWG}$, as a function of time during a single calibration run.
        (Bottom panel) The demodulated response of a single detector.
        We use the $Q$ and $U$ response only when the SWG is kept at each measurement angle for 10 seconds.
        % The SWG was kept at each measurement angle for 10 seconds.
    }
    \label{fig:tod_during_wiregrid_cal}
\end{figure}

Since the signal is modulated at 4 times the HWP rotation frequency, the SWG polarized signal appears as a $4f$-harmonic, as shown in Figure~\ref{fig:calibration_operation}.
The relationship between $\theta_\mathrm{det}$ and $\theta^{(j)}_\mathrm{SWG}$ is obtained from demodulation by using the encoder data of the HWP rotation.
The TOD are demodulated by applying a band-pass filter, $\mathcal{F}_\mathrm{BP}$, deconvolving the HWP fast-axis response,
and then followed by a low-pass filter, $\mathcal{F}_\mathrm{LP}$. The demodulated signal is
\begin{equation}
    \begin{aligned}
    d_\mathrm{demod}^{(j)} & = \mathcal{F}_\mathrm{LP}\!\left[~e^{i4\omega t}\; \mathcal{F}_\mathrm{BP}\!\left\{d^{(j)}_\mathrm{mod}(t)\right\}\right] \\
    & = \left(A_\mathrm{SWG} \exp{2i\theta^{(j)}_\mathrm{SWG}} + \bm{\varepsilon}\right)e^{2i\theta_\mathrm{det} + 4i\omega\tau_\mathrm{det}}\\
    & \equiv Q^{(j)} + \varepsilon_Q + i (U^{(j)} + \varepsilon_U).
    \end{aligned}
    \label{eq:demod}
\end{equation}
$\mathcal{F}_\mathrm{BP}$ targets the $4f$-harmonic with a passband of 7.62--8.85~Hz (shaded range in Figure~\ref{fig:calibration_operation}).
The real and imaginary parts correspond to the Stokes parameters $Q$ and $U$, respectively.
We extract $Q^{(j)}$ and $U^{(j)}$ at each measurement angle of the SWG rotation.
Figure \ref{fig:tod_during_wiregrid_cal} shows the direction of the SWG and the demodulated response as a function of time in a single calibration run.
The observed polarization response clearly tracks the stepwise rotation of the SWG.

\begin{figure*}[t!]
    \centering
    \includegraphics[width=0.9\textwidth]{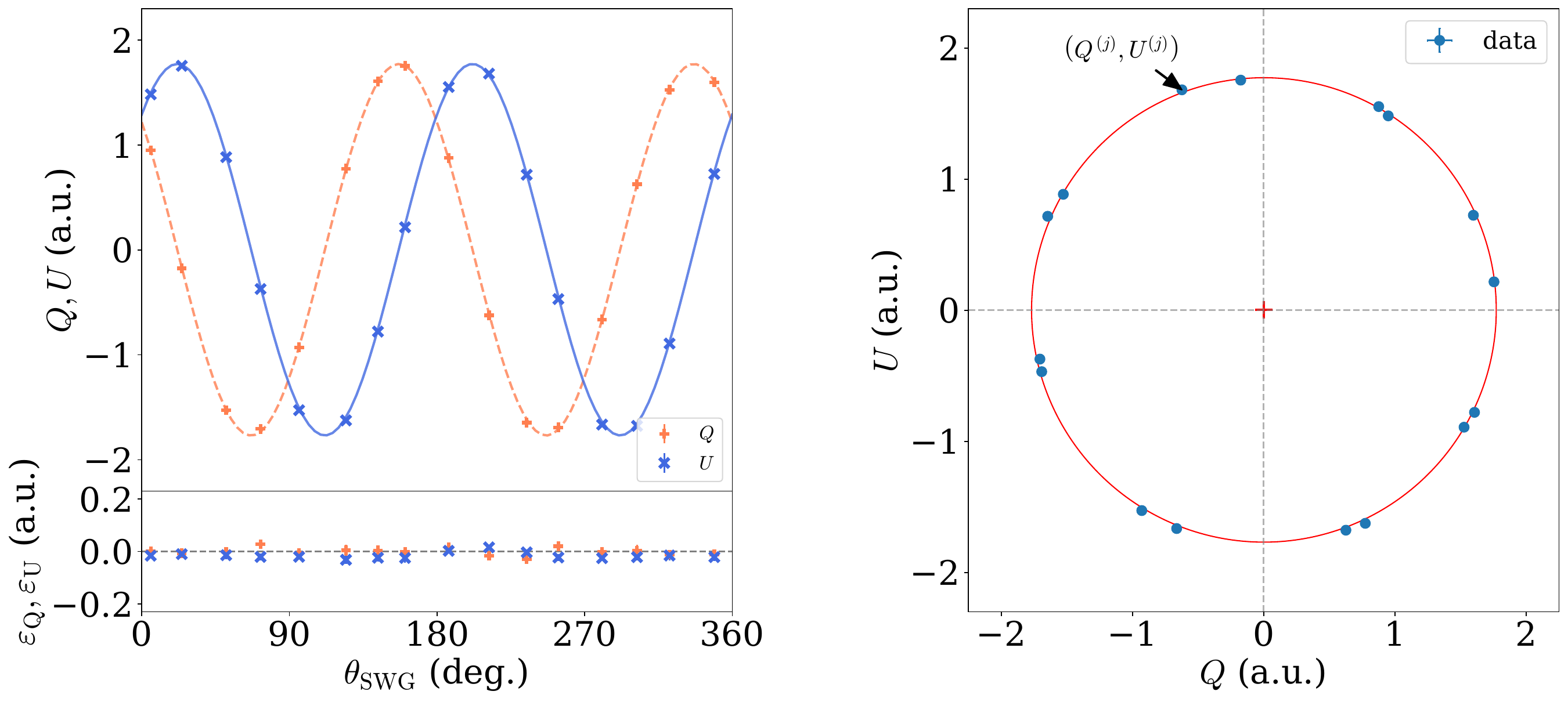}
    \caption{
    (Left panel) The demodulated signal, $Q$ and $U$, as a function of the SWG angle.
    The lower panel shows the residuals from the curves with the best fit parameters.
    They are about two orders of magnitude smaller than the wire grid signal.
    (Right panel) Demodulated data points on the $Q$--$U$ plane, i.e., there are 16 pairs of $Q$ and $U$ corresponding to each measurement angle of the SWG rotation.
    The circle shown here is the one best fitting the data points in the $Q$--$U$ plane.
    The obtained results of this detector are $A_\mathrm{SWG} = 1.774\pm 0.002$, $\varepsilon_Q = (0.007\pm 0.292)\times 10^{-2}$, and $\varepsilon_U = (-0.316\pm 0.316)\times 10^{-2}$.
    }
    \label{fig:qu_to_wires_and_cal_circle}
\end{figure*}

\begin{figure}[t!]
    \centering
    \includegraphics[width=0.8\columnwidth]{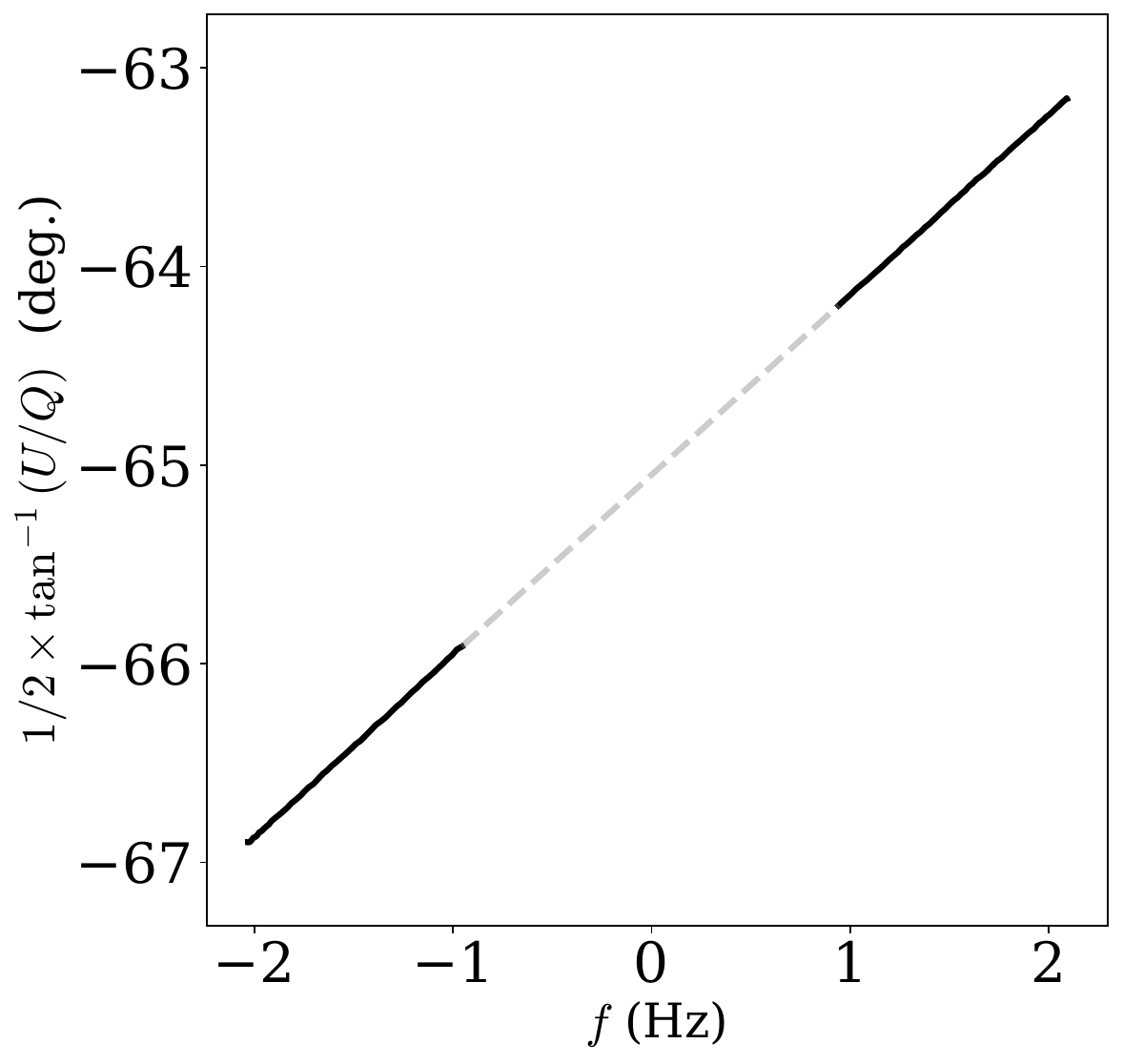}
    \caption{
        An extracted angle as a function of the HWP rotation frequency, $f$, for a single detector.
        We can stably demodulate the TOD within the range $|f| \geq 1~\mathrm{Hz}$.
        We extract the time constant of this detector by fitting the data with Equation \ref{eq:det_hat}.
    }
    \label{fig:time_const_operation}
\end{figure}

\begin{figure*}[t!]
    \centering
    \includegraphics[width=0.95\textwidth]{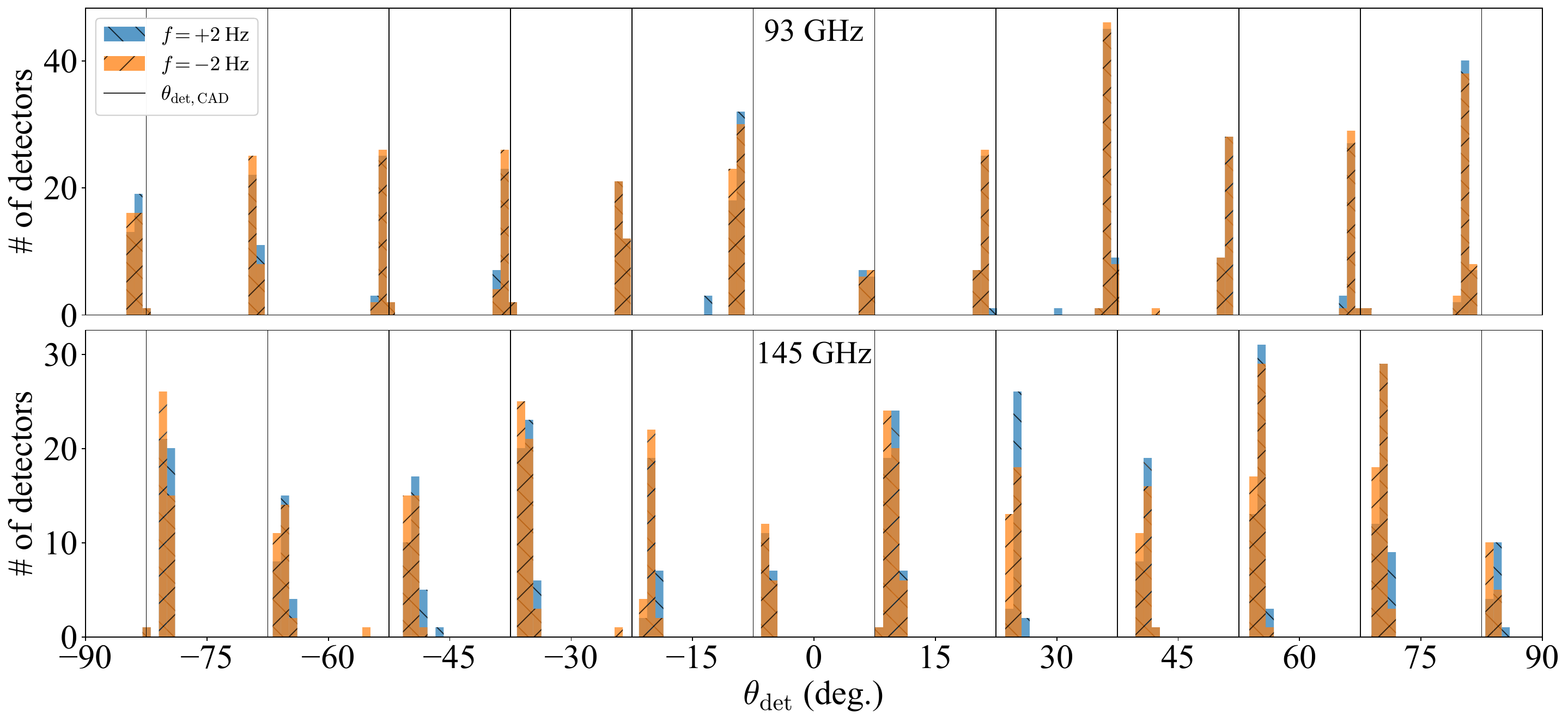}
    \caption{
        Distribution of the calibrated detector polarization angles for the 93~GHz (top) and 145~GHz (bottom) bands.
        There are 12 distinct populations corresponding to the 12 polarization sensitive directions defined in the detector module design.
        Two types of plots are shown: the $\theta_\mathrm{det}$ from the clockwise HWP rotation ($f=+2~\mathrm{Hz}$, blue) and the counterclockwise HWP rotation ($f=-2~\mathrm{Hz}$, orange).
        Both HWP rotation directions produce consistent detector polarization angles.
        Dotted lines indicate the design angles, $\theta_{\mathrm{det, CAD}}$, expected from the CAD specification adapted for the receiver assembly.
        The difference between 93~GHz and 145~GHz is consistent with the expected phase shift due to the achromaticity of the HWP \citep{SOHWPDesign}.
    }
    \label{fig:calibrated_angle}
\end{figure*}

\begin{figure*}[t!]
    \centering
    \includegraphics[width=1.90\columnwidth]{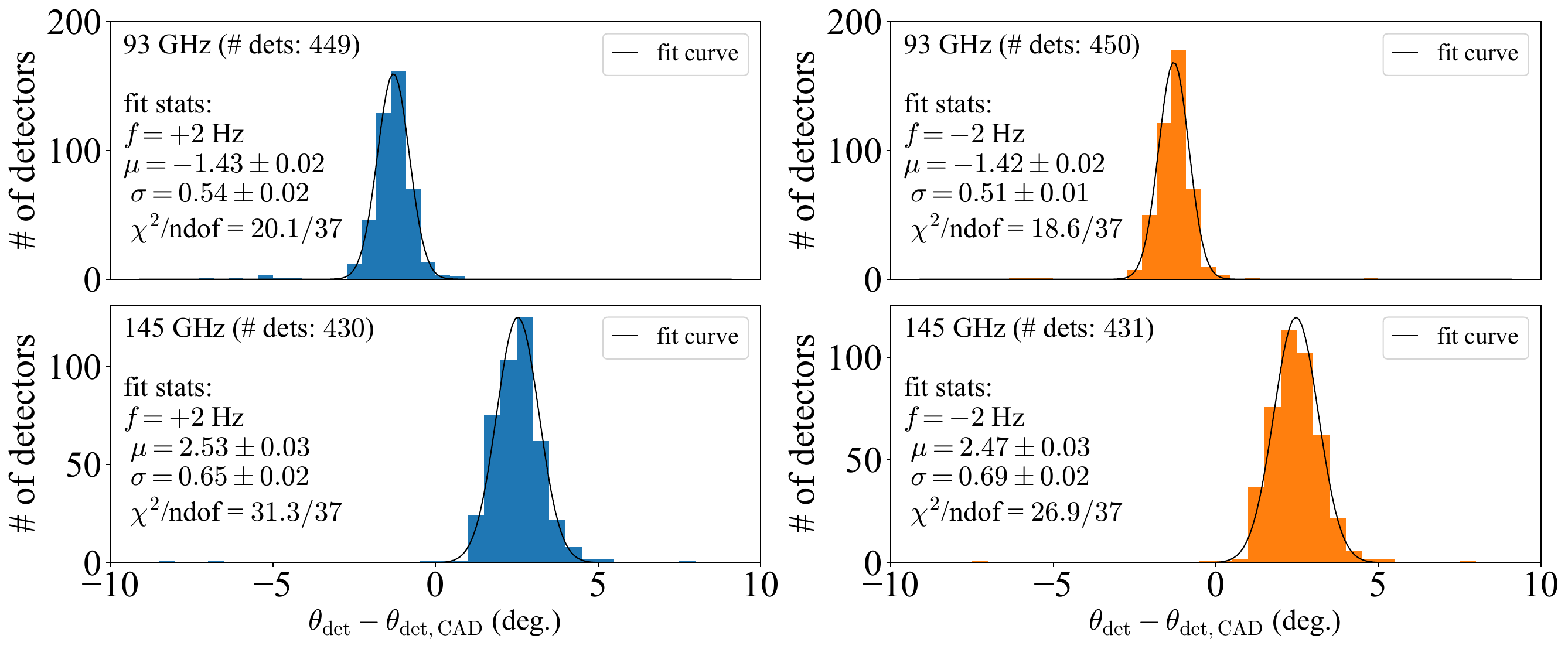}
    \caption{
        Distribution of the calibrated detector polarization angles with respect to the design angles for 93~GHz (top) and 145~GHz (bottom).
        The left (right) panels correspond to the data taken with the clockwise (counterclockwise) HWP rotation.
        The estimated mean, $\mu$, and standard deviation, $\sigma$, and the goodness of fit, $\chi^2/\mathrm{ndof}$, from the Gaussian fits to all 40 angular bins are also shown.
        The clustering of most detectors in single locations for each panel demonstrates the successful operation of the detector identification algorithm in SO \citep{Lashner_2024}. This algorithm determines the correct mapping between each readout channel and the physical detector on the focal plane.
    }
    \label{fig:single_distribution}
\end{figure*}

\subsection{Analysis for the SWG signal} \label{subsec:calibration}

Figure~\ref{fig:qu_to_wires_and_cal_circle} shows the demodulated Stokes parameters, $Q$ and $U$, as a function of the SWG angle, and the relations of $Q$ and $U$ at each measurement angle of the SWG.
We extract $A_\mathrm{SWG}$ and the polarization offsets $(\varepsilon_Q, \varepsilon_U)$ for each detector by fitting a circle in the $Q$--$U$ plane to data from the 16 SWG measurement angles for the first and second data sets.
We confirmed that the offset is approximately two orders of magnitude smaller than $A_\mathrm{SWG}$ as shown in Figure~\ref{fig:qu_to_wires_and_cal_circle}.
The measured offsets are consistent with zero.

By averaging over all the 16 measurement angles, the detector polarization angle (without the effect of $\tau_\mathrm{det}$) is extracted as
\begin{equation}
  \hat{\theta}_\mathrm{det}
  \equiv \frac{1}{N} \sum_{j=1}^N
    \left[
      \frac{1}{2} \tan^{-1}
        \left(
          \frac{U^{(j)} - \varepsilon_U}
               {Q^{(j)} - \varepsilon_Q}
        \right)
      - \theta^{(j)}_\mathrm{SWG}
    \right],
  \label{eq:det}
\end{equation}
where $N=16$. The uncertainty in $\hat{\theta}_\mathrm{det}$ is given by the the root-mean-square deviations of the residuals of the best-fitting circle divided by $\sqrt{2N}A_\mathrm{SWG}$, which is typically $0.06^\circ$.
We obtain the true detector polarization angle, $\theta_\mathrm{det}$, with the following correction:
\begin{equation}
    \hat{\theta}_\mathrm{det} = \theta_\mathrm{det} + 4\pi f\tau_\mathrm{det}.
    \label{eq:det_hat}
\end{equation}
The procedure to obtain the $\tau_\mathrm{det}$ estimate is presented in the next section.

\subsection{Time constant estimation} \label{subsec:tc}

The time constant can be obtained from the SWG data by changing the rotation velocity of the HWP \citep{Simon_2014}.
Figure~\ref{fig:time_const_operation} shows the estimated polarization angle of a single detector as a function of the HWP rotation frequency.
The demodulation band-pass filter for the time constant estimation was set to 4.0--8.0~Hz to select stably modulated data.
The slope of the line is $4\pi\tau_\mathrm{det}$.
The median measured time constants are $2.06\pm 0.10 ~\mathrm{ms}~(93~\mathrm{GHz})$ and $1.13\pm 0.05~\mathrm{ms}~(145~\mathrm{GHz})$.

\section{Results and Discussion} \label{sec:evaluation}

The detector polarization angles, $\theta_\mathrm{det}$, are obtained from $\hat{\theta}_\mathrm{det}$ and $\tau_\mathrm{det}$ calculated from Equations~\ref{eq:det} and \ref{eq:det_hat}.
Figure~\ref{fig:calibrated_angle} shows the distributions of the detector polarization angles for the 93~GHz and 145~GHz bands, respectively.
We observed 12 distinct populations of detector polarization angles in our measurements which correspond to the 12 distinct polarization-sensitive orientations defined in the detector module design \citep{AdvACT}.
We obtained consistent results between $\theta_\mathrm{det}$ derived from the data taken with both counterclockwise ($+2~\mathrm{Hz}$) and clockwise ($-2~\mathrm{Hz}$) HWP rotation.

We obtain distributions of residual detector offsets by subtracting the design angles ($\theta_\mathrm{det,\,CAD}$) from the measured angles ($\theta_\mathrm{det}$) for each band and direction of the HWP rotation, as shown in Figure~\ref{fig:single_distribution}.
The calibrated detector polarization angles have a non-zero angle offset, which are arbitrary offsets common to all detectors.
Future work will correct for the projection for the SWG orientation on the sky, determined using the tilt sensor as a gravity reference.

The histograms in Figure~\ref{fig:single_distribution} are well described by Gaussian fits, confirming that the detectors respond coherently to the artificial polarization signal generated by the SWG.
The statistical uncertainties on the mean of the distributions are $0.02^\circ$ for the 93~GHz band and $0.03^\circ$ for the 145~GHz band.
They indicate the statistical precision of the global polarization angle for the single detector module, and they are better than our requirement.

We neglect the time variation of $\varepsilon_Q$ and $\varepsilon_U$ when we measure the time constant in Section~\ref{subsec:tc}.
Its impact for the detector polarization angle is $0.01^\circ$, which is small compared to our requirement. % ($<0.2^\circ$).

The difference between the 93~GHz and 145~GHz bands ($3.92^\circ\pm0.04^\circ$) is explained by an expected phase shift due to the achromaticity of the 3-layer-stacked HWP \citep{SOHWPDesign}.
The expected value is calculated as twice the phase shift of the HWP based on Equation \ref{eq:theta}.
It is $\simeq 4^\circ$ when the frequency-dependent phase \citep{SOHWPDesign} is averaged over the detector passband \citep{Sierra_2025}.
Future detector passband measurements using a fourier transform spectrometer will confirm this explanation with better precision.
All global rotation offsets, including the frequency-dependent phase shift of the achromatic HWP, will be accounted for by use of the gravity reference to project the SWG orientation onto the sky.
These effects are not critical for the purpose of the per-frequency-band relative calibration.

\section{Conclusion} \label{sec:conclusion}

We demonstrate a fully remote polarization method using a sparse wire grid.
The single calibration, which includes the detector time-stream acquisition, demodulation of the HWP, and definition relative to the SWG is working correctly.
We observed the $Q-U$ responses clearly forming a circle in the demodulated data.
The signal and modeling of the SWG polarization is within expectation, and measurements yield statistical uncertainties of the calibrated detector polarization are $0.02^\circ$ and $0.03^\circ$ at 93~GHz and 145~GHz, respectively.
The systematic uncertainty from the hardware fabrication was evaluated to be $0.08^\circ$ in our previous study \citep{SOWireGridDesign}.
Their sum in quadrature is $0.09^\circ$.

In this study, we demonstrate the relative angle calibration.
Future work will incorporate data from the tilt sensor equipped on the SWG system.
The tilt sensor provides the gravity reference, and it allows us to determine the absolute orientation of the SWG.
We will also extend this analysis to all seven detector modules in the SAT.
Furthermore, comparison with other calibration methods, such as observations of Tau\,A, is important for future studies.

\section{Acknowledgement} \label{sec:acknowledge}

This work was supported in part by a grant from the Simons Foundation (Award \#457687, B.K.).
This work was supported by the U.S. National Science Foundation (Award Number: 2153201).
This work was also supported by JSPS KAKENHI Grant Numbers
JP17H06134, JP18J01039, JP19H00674, JP22H04913, JP23H00105, JP24KJ1333, JP24K23938, and the JSPS Core-to-Core Program JPJSCCA20200003.
HN acknowledges research support program for JSPS-Fellows in cooperation with European Research Council (ERC).
SA acknowledges support from the Foundation of Kinoshita Memorial Enterprise.
SG acknowledges support from STFC and UKRI (grant numbers ST/W002892/1 and ST/X006360/1).
FN acknowledges funding from the European Union (ERC, POLOCALC, 101096035).

\bibliography{ref}{}
\bibliographystyle{aasjournal}

\end{document}